\documentclass[journal,epsf]{IEEEtran}
\usepackage{amsmath}
\usepackage{amsfonts}
\usepackage{amssymb}
\usepackage{graphicx}

\ifCLASSINFOpdf
\else
\fi
\newtheorem{theorem}{Theorem}
\newcommand {\dfn} {\stackrel{\Delta} {=}}

\newcommand {\reals} {{\rm I\!R}}

\newcommand {\hs} {\hat{s}}

\newcommand {\bx} {\mbox{\boldmath $x$}}
\newcommand {\by} {\mbox{\boldmath $y$}}

\newcommand {\bE} {\mbox{\boldmath $E$}}

\newcommand {\bY} {\mbox{\boldmath $Y$}}

\newcommand{\calE}{{\cal E}}

\newcommand{\calN}{{\cal N}}

\newcommand{\calX}{{\cal X}}
\newcommand{\calY}{{\cal Y}}

\begin{document}
\title{Rate--Distortion Function via Minimum Mean Square
Error Estimation}
\author{Neri Merhav
\thanks{N.~Merhav is with the Department
of Electrical Engineering, Technion -- Israel Institute of Technology, Haifa,
32000, Israel. E-mail: merhav@ee.technion.ac.il.}}

\maketitle

\begin{abstract}
\boldmath
We derive a simple general
parametric representation of the rate--distortion function
of a memoryless source,
where both the rate and the distortion 
are given by integrals whose integrands include
the minimum mean square error (MMSE) of the distortion $\Delta=d(X,Y)$ based on the
source symbol $X$, with respect to a certain joint distribution of these two random
variables. At first glance, these relations may seem somewhat similar to the I--MMSE
relations due to Guo, Shamai and Verd\'u, but they are, in fact, quite
different. The new relations among rate, distortion, and MMSE
are discussed from several aspects, and more importantly, it is demonstrated 
that they can sometimes be 
rather useful for obtaining non--trivial upper and lower bounds on the
rate--distortion function, as well as for determining the
exact asymptotic behavior for very low and for very large distortion.
Analogous MMSE relations hold for channel capacity as well.
\end{abstract}
\IEEEpeerreviewmaketitle

\section{Introduction}

\IEEEPARstart{I}{t} has been
well known for many years that the derivation of the 
rate--distortion function of a given source and distortion measure,
does not lend itself to closed form expressions, even
in the memoryless case, except for
a few very simple examples
\cite{Berger71},\cite{CT06},\cite{CK81},\cite{Gray90}.
This has triggered the derivation of some upper and lower bounds,
both for memoryless sources and for sources with memory.

One of the most important lower bounds on the rate--distortion function,
which is applicable for difference distortion measures (i.e., distortion
functions that depend on their two arguments only through the difference between
them), is the Shannon lower bound in its different forms, e.g., the discrete
Shannon lower bound, the continuous Shannon lower bound, and the 
vector Shannon lower bound. This family of bounds is especially
useful for semi-norm--based distortion measures \cite[Section 4.8]{Gray90}.
The Wyner--Ziv lower bound 
\cite{WZ71} for a source with memory is a
convenient bound, which is based on the rate--distortion function 
of the memoryless source formed from the product measure pertaining to the single--letter
marginal distribution of the original source and it may be combined elegantly
with the Shannon lower bound.
The autoregressive lower bound
asserts that the
rate--distortion function 
of an autoregressive source
is lower bounded by the rate--distortion function
of its innovation process, which is again, a memoryless source.

Upper bounds are conceptually easier to derive, as they may result from the
performance analysis of a concrete coding scheme, or from 
random coding with respect to
(w.r.t.) an arbitrary random coding distribution, etc.
One well known example is the Gaussian upper bound, which upper bounds the
rate--distortion 
function of an arbitrary memoryless (zero--mean) source w.r.t.\ the
squared error distortion measure by the rate--distortion
function of the Gaussian source with the same second moment.
If the original source has memory, then the same principle generalizes
with the corresponding Gaussian source having the same autocorrelation function
as the original source \cite[Section 4.6]{Berger71}. 

In this paper, we focus on
a simple general
parametric representation of the rate--distortion function
which seems to set the stage for the derivation 
of a rather wide family of both 
upper bounds and lower bounds on the rate--distortion function.
In this parametric representation, both the rate and the distortion
are given by integrals whose integrands include
the minimum mean square error (MMSE) of the distortion based
on the source symbol, with respect to a certain joint distribution of these two
random variables. More concretely, given a memoryless source designated by a random
variable (RV) $X$, governed by a probability function\footnote{Here, and
throughout the sequel,
the term ``probability function'' refers to a probability mass function
in the discrete case and to a probability density function in
the continuous case.} $p(x)$, 
a reproduction variable $Y$, governed by a probability function $q(y)$,
and a distortion measure $d(x,y)$, the rate and the distortion can be
represented parametrically via a real parameter $s\in[0,\infty)$ as follows:
\begin{eqnarray}
\label{distortion}
D_s&=&D_0-\int_0^s\mbox{d}\hs\cdot\mbox{mmse}_{\hs}(\Delta|X)\nonumber\\
&=&D_\infty+\int_s^\infty\mbox{d}\hs\cdot\mbox{mmse}_{\hs}(\Delta|X)
\end{eqnarray}
and
\begin{eqnarray}
\label{rate}
R_q(D_s)&=&\int_0^s\mbox{d}\hs\cdot\hs\cdot\mbox{mmse}_{\hs}(\Delta|X)\nonumber\\
&=&R_q(D_\infty)-\int_s^\infty\mbox{d}\hs\cdot\hs\cdot\mbox{mmse}_{\hs}(\Delta|X),
\end{eqnarray}
where $D_s$ is the distortion pertaining to parameter value $s$,
$R_q(D_s)$ is the rate--distortion function w.r.t.\ reproduction
distribution $q$, computed at $D_s$, 
$\Delta=d(X,Y)$, and $\mbox{mmse}_s(\Delta|X)$ is the MMSE of estimating
$\Delta$ based on $X$, where the joint probability function of $(X,\Delta)$ is induced
by the following joint probability function of $(X,Y)$:
\begin{equation}
p_s(x,y)=p(x)\cdot w_s(y|x)=p(x)\cdot\frac{q(y)e^{-sd(x,y)}}{Z_x(s)}
\end{equation}
where $Z_x(s)$ is a normalization constant, given by
$\int\mbox{d}yq(y)e^{-sd(x,y)}$ in the continuous case, or
$\sum_yq(y)e^{-sd(x,y)}$ in the discrete case.

At first glance, eq.\ (\ref{rate}) looks somewhat similar to the
I--MMSE relation of \cite{GSV05},
which relates the mutual information between the input and the output
of an additive white Gaussian noise (AWGN) 
channel and the MMSE of estimating the channel input based
on the noisy channel output. 
As we discuss later on, however,
eq.\ (\ref{rate}) is actually very different from the I-MMSE relation
in many respects. In this context, it is important to emphasize that
a relation analogous to (\ref{rate}) applies also to channel capacity,
as will be discussed in the sequel.

The relations (\ref{distortion}) and 
(\ref{rate}) have actually already been raised in a
companion paper \cite{Merhav09a}
(see also \cite{Merhav09b} for a conference version). 
Their derivation there was triggered and inspired by certain analogies 
between the rate--distortion problem and statistical mechanics, which were
the main theme of that work. However, the significance and the usefulness
of these rate--distortion-MMSE 
relations were not explored in \cite{Merhav09a} and \cite{Merhav09b}.

It is the purpose of the present work to study these
relations more closely and to demonstrate their utility,
which is, as said before, in deriving upper and lower bounds.
The underlying idea is that bounds on $R_q(D)$
(and sometimes also on $R(D)=\min_qR_q(D)$)
may be obtained via relatively simple bounds on
the MMSE of $\Delta$ based on $X$. These bounds can either
be simple technical bounds
on the expression of the MMSE itself, or bounds that stem
from pure estimation--theoretic considerations. For example, upper bounds 
may be derived by analyzing the MMSE of a certain sub-optimum estimator, e.g., 
a linear estimator, which is easy to analyze.
Lower bounds can be taken from the available plethora of lower bounds
offered by estimation theory, e.g., the Cram\'er--Rao lower bound.

Indeed, an important part of this work is a section of examples,
where it is demonstrated how to use the proposed relations and
derive explicit bounds from them. In one of these 
examples, we derive two sets of
upper and lower bounds, one for a certain range of low distortions and
the other, for high distortion values. At both edge-points of the 
interval of distortion values of interest, the corresponding upper and
lower bound asymptotically approach the limiting value with the same
leading term, and so, they sandwich the exact asymptotic behavior of the
rate--distortion function, both in the low distortion limit and in the
high distortion limit.

The outline of this paper is as follows. In Section II,
we establish notation conventions. In Section III,
we formally present the main result, prove it, and discuss its
significance from the above--mentioned aspects. In Section IV, we
provide a few examples that demonstrate the usefulness of the MMSE
relations. Finally, in Section V, we summarize and conclude.

\section{Notation Conventions}
\label{notation}

Throughout this paper,
RV's will be denoted by capital
letters, their sample values will be denoted by
the respective lower case letters, and their alphabets will be denoted
by the respective calligraphic letters.
For example, $X$ is a random variable, $x$ is a specific realization
of $X$, and $\calX$ is the alphabet in which $X$ and $x$ take on values.
This alphabet may be finite, countably infinite, or a continuum, like the
real line $\reals$ or an interval $[a,b]\subset\reals$.

Sources and channels will be denoted generically by the letter $p$, or $q$,
which will designate also their corresponding probability functions, i.e.,
a probability density function (pdf) in the continuous case, or a probability mass
function (pmf) in the discrete case.
Information--theoretic quantities, like entropies and mutual
informations, will be denoted according to the usual conventions
of the information theory literature, e.g., $H(X)$, $I(X;Y)$,
and so on. If a RV is continuous--valued, then its differential entropy
and conditional differential entropy
will be denoted with $h$ instead of $H$, i.e., $h(X)$ is the conditional
differential entropy of $X$, $h(X|Y)$ is the conditional differential entropy of $X$
given $Y$, and so on. The expectation operator will be denoted, as usual,
by $\bE\{\cdot\}$.

Given a source RV $X$, governed by a probability function $p(x)$, $x\in\calX$,
a reproduction RV $Y$, governed by a probability function $q(y)$, $y\in\calY$,
and a distortion measure $d:\calX\times\calY\to \reals^+$, we define the
rate--distortion function of $X$ w.r.t.\ distortion measure $d$ and
reproduction distribution $q$ as
\begin{equation}
R_q(D)\dfn\min I(X;Y),
\end{equation}
where $X\sim p$ and the minimum is across 
all channels $\{w(y|x),~x\in\calX,~y\in\calY\}$
that satisfy $\bE\{d(X,Y)\}\le D$ and $\bE\{w(y|X)\}=q(y)$ for all
$y\in\calY$. Clearly, the rate--distortion function, $R(D)$, is given
by $R(D)=\inf_qR_q(D)$. We will also use the notation $\Delta\dfn d(X,Y)$.
Obviously, since $X$ and $Y$ are RV's, then so is $\Delta$.

\section{MMSE Relations: Basic Result and Discussion}

Throughout this section, our definitions will assume that both $\calX$ and
$\calY$ are finite alphabets. Extensions to continuous alphabets will be obtained
by a limit of fine quantizations, with summations eventually being replaced by integrations.

Referring to the notation defined in Section \ref{notation}, 
for a given positive real $s$, define the conditional probability function
\begin{equation}
w_s(y|x)\dfn\frac{q(y)e^{-sd(x,y)}}{Z_x(s)}
\end{equation}
where
\begin{equation}
Z_x(s)\dfn\sum_{y\in\calY}q(y)e^{-sd(x,y)}
\end{equation}
and the joint pmf
\begin{equation}
p_s(x,y)=p(x)w_s(y|x).
\end{equation}
Further, let
\begin{eqnarray}
\mbox{mmse}_s(\Delta|X)&=&\bE_s\{[\Delta-\bE\{\Delta|X\}]^2\}\nonumber\\
&=&\bE_s\{[d(X,Y)-\bE_s\{d(X,Y)|X\}]^2\}
\end{eqnarray}
where $\bE_s\{\cdot\}$ is the expectation operator w.r.t.\ $\{p_s(x,y)\}$,
and defining $\psi(x)$ as the conditional expectation $\bE_s\{d(x,Y)|X=x\}$
w.r.t.\ $\{w_s(y|x)\}$, $\bE_s\{d(X,Y)|X\}$ is defined as $\psi(X)$.

Our main result, in this section, is the following (the proof appears in the
Appendix):

\noindent
\begin{theorem}
\label{thm1}
The function $R_q(D)$ can be represented parametrically via the parameter
$s\in[0,\infty)$ as follows: 
\begin{itemize}
\item[(a)]
The distortion is obtained by
\begin{eqnarray}
\label{dist1}
D_s&=&D_0-\int_0^s\mbox{d}\hs\cdot\mbox{mmse}_{\hs}(\Delta|X)\nonumber\\
&=&D_\infty+\int_s^\infty\mbox{d}\hs\cdot\mbox{mmse}_{\hs}(\Delta|X)
\end{eqnarray}
where 
\begin{equation}
D_0=\sum_{x,y}p(x)q(y)d(x,y) 
\end{equation}
and 
\begin{equation}
D_{\infty}=\sum_xp(x)\min_yd(x,y).
\end{equation}
\item[(b)]
The rate is given by
\begin{eqnarray}
\label{rate1}
& &R_q(D_s)\nonumber\\
&=&\int_0^s\mbox{d}\hs\cdot\hs\cdot\mbox{mmse}_{\hs}(\Delta|X)\nonumber\\
&=&R_q(D_\infty)-\int_s^\infty\mbox{d}\hs\cdot\hs\cdot\mbox{mmse}_{\hs}(\Delta|X).
\end{eqnarray}
\end{itemize}
\end{theorem}

In the remaining part of this section, we discuss the significance
and the implications of Theorem 1 from several aspects.\\

\noindent
{\it Some General Technical Comments}\\

The parameter $s$ has the geometric meaning of the negative local slope
of the function $R_q(D)$. This is easily seen by taking the derivatives
of (\ref{dist1}) and (\ref{rate1}), i.e.,
$\mbox{d}R_q(D_s)/\mbox{d}s=s\cdot\mbox{mmse}_s(\Delta|X)$ and
$\mbox{d}D_s/\mbox{d}s=-\mbox{mmse}_s(\Delta|X)$, whose ratio is $R_q'(D_s)=-s$.
This means also that the parameter $s$ 
plays the same role as in the well known parametric representations
of \cite{Berger71} and \cite{Gray90}, which is to say that it can also be thought
of as the Lagrange multiplier of the minimization of
$[I(X;Y)+s\bE\{d(X,Y)\}]$ subject to the reproduction distribution constraint.

On a related note, we point out that 
Theorem \ref{thm1} is based on the following representation of
$R_q(D)$:
\begin{equation}
\label{lgd1}
R_q(D)=-\min_{s\ge 0}\left[sD+\sum_{x\in\calX}p(x)\ln Z_x(s)\right],
\end{equation}
which we prove in the Appendix as the first step in the proof of Theorem 1.

It should be emphasized that the pmf $q$, that plays a role in the
definition of $w_{\hs}(y|x)$ (and hence also the definition of
$\mbox{mmse}_{\hs}(\Delta|X)$) should be kept {\it fixed} throughout the
integration, independently of the integration variable $\hs$, since it is the same
pmf as in the definition of $R_q(D)$. Thus, even if $q$ is known to
be optimum for a given target distortion $D$ (and then it yields $R(D)$), the pmf $q$
must be kept unaltered throughout the integration, in spite of the fact that
for other values of $\hs$ (which correspond to other distortion levels),
the optimum reproduction pmf might be different. In particular, note
that the marginal of $Y$, that is induced from the joint pmf $p_s(x,y)$, may not
necessarily agree with $q$. Thus, $p_{\hs}(x,y)$ should only be considered as an
auxiliary joint distribution that defines $\mbox{mmse}_{\hs}(\Delta|X)$.\\

\noindent
{\it Using Theorem 1 for Bounds on $R_q(D)$}\\

As was briefly explained in the Introduction (and 
will also be demonstrated in the next section), Theorem \ref{thm1}
may set the stage for the derivation of upper and lower bounds
to $R_q(D)$ for a general 
reproduction distribution $q$, and hence also for the
rate--distortion function $R(D)$ when the optimum $q$ is happened
to be known or is easily derivable 
(e.g., from symmetry and convexity considerations). 

The basic underlying
idea is that bounds on $R_q(D)$ may be induced from bounds on 
$\mbox{mmse}_{\hs}(\Delta|X)$ across the integration interval. The bounds
on the MMSE may either be derived from purely technical considerations,
upon analyzing the expression of the MMSE directly, or by using
estimation--theoretic tools. In the latter case, lower bounds may be
obtained from fundamental lower bounds to the MMSE, like the Bayesian
Cram\'er--Rao bound, or more advanced lower bounds available 
from the estimation theory literature, for example,
the Weiss--Weinstein bound 
\cite{WW85},\cite{Weiss85}, whenever applicable. Upper bounds may be
obtained by analyzing the mean square error (MSE) of a specific
(sub-optimum) estimator, which is relatively easy to analyze, or more
generally by analyzing the performance of the best estimator within
a certain limited class of estimators, like the class of linear estimators
of the `observation' $X$, or a certain fixed function of $X$.

In Theorem 1 we have deliberately presented two integral forms for
both the rate and the distortion. As $D_s$ is
monotonically decreasing and $R_q(D_s)$ is 
monotonically increasing in $s$, the integrals at the first lines of
both eqs.\ (\ref{dist1}) and (\ref{rate1}),
which include relatively small values of $\hs$, 
naturally lend themselves to
derivation of bounds in the low--rate (high distortion) regime, whereas the
second lines of these equations are more suitable in low--distortion
(high resolution) region. For example, to derive an upper 
bound on $R_q(D)$ in the high--distortion range, one would need
a lower bound on $\mbox{mmse}_{\hs}(\Delta|X)$ to be used in the
first line of (\ref{dist1}) and
an upper bound on $\mbox{mmse}_{\hs}(\Delta|X)$ to be substituted into the
first line of (\ref{rate1}). If one can then derive, from the former, an upper bound on
$s$ as a function of $D$, and substitute it into the upper bound on
the rate in terms on $s$,
then this will result in an upper bound to $R_q(D)$.
A similar kind of reasoning is applicable to 
the derivation of other types of bounds. This point will be demonstrated
mainly in Examples C and D in the next section.\\

\noindent
{\it Comparison to the I--MMSE Relations}\\

In the more conceptual level, 
item (b) of Theorem 1 may remind the familiar reader
about well--known results due to Guo, Shamai and Verd\'u \cite{GSV05},
which are referred to as I--MMSE relations (as well as later works
that generalize these relations). The similarity between eq.\ (\ref{rate1})
and the I--MMSE relation (in its basic form)
is that in both cases a mutual information
is expressed as an integral whose integrand includes the MMSE of a certain
random variable (or vector) given some observation(s). However, to the
best of our judgment, this is the only similarity. 

In order to sharpen
the comparison between the two relations, it is instructive to look at
the special case where all random variables are Gaussian and the distortion
measure is quadratic: In the 
context of Theorem 1, consider $Y$ to be a zero--mean Gaussian RV with 
variance $\sigma_y^2$,
and let $d(x,y)=(x-y)^2$. As will be seen in Example B of the next section,
this then means that $w_s(y|x)$ can be described by the additive Gaussian
channel $Y=aX+Z$, where $a=2s\sigma_y^2/(1+2s\sigma_y^2)$ and $Z$ is a
zero--mean Gaussian RV, independent of $X$, and with variance
$\sigma_y^2/(1+2s\sigma_y^2)$. Here, we have $\Delta=(Y-X)^2=[Z-(1-a)X]^2$.
Thus, the integrand of (\ref{rate1}) includes the MMSE in estimating
$[Z-(1-a)X]^2$ based on the {\it channel input} $X$. It is therefore about
estimating a certain 
function of $Z$ and $X$, where $X$ is the observation at hand and
$Z$ is independent of $X$. 

This is very different from the paradigm of the
I--MMSE relation: there the channel is $Y=\sqrt{\mbox{\sl snr}}X+Z$, where $Z$ is
standard normal, the integration variable is $\mbox{\sl snr}$, and the estimated
RV is $X$ (or equivalently, $Z$) based on the {\it channel output}, $Y$. Also,
by comparing the two channels, it is readily seen that the
integration variable $s$, in our setting, can be related to the integration
variable, $\mbox{\sl snr}$, of the I-MMSE relation
according to 
\begin{equation}
\mbox{\sl snr}=\frac{4s^2}{\sigma_y^2(1+2s\sigma_y^2)},
\end{equation}
and so, the relation between the two integration variables is highly
non--linear. We therefore observe that the two MMSE results are fairly different.\\

\noindent
{\it Analogous MMSE Formula for Channel Capacity}\\

Eq.\ (\ref{lgd1}) can be understood conveniently 
as an achievable rate using a simple random
coding argument (see Appendix): 
The coding rate $R$ should be (slightly larger than) the large
deviations rate
function of the probability of the event $\{\sum_{i=1}^n
d(x_i,Y_i)\le nD\}$, where $(x_1,\ldots,x_n)$ is a typical source sequence
and $(Y_1,\ldots,Y_n)$
are drawn i.i.d.\ from $q$. As is well known, a similar random coding argument applies to
channel coding (see also \cite{Merhav08}): Channel capacity can be obtained as
the large deviations rate function of
the event $\{\sum_{i=1}^n
d(X_i,y_i)\le nD\}$, where
now $(y_1,\ldots,y_n)$ is a channel output sequence typical to $q$,
$(X_1,\ldots,X_n)$ are drawn i.i.d.\ according to a given
input pmf $\{p(x)\}$, the distortion measure is
chosen to be $d(x,y)=-\ln w(y|x)$ ($\{w(y|x)\}$ being the channel transition
probabilities) and $D=H(Y|X)$. Thus, the analogue of (\ref{lgd1}) is
\begin{equation}
C_p=-\min_{s\ge 0}\left[sH(Y|X)+\sum_{y\in
Y}q(y)\ln Z_y(s)\right]
\end{equation}
where
\begin{equation}
Z_y(s)=\sum_{x\in\calX}p(x)w^s(y|x)
\end{equation}
and the minimizing $s$ is always $s^*=1$.
Consequently, the corresponding integrated MMSE formula would read
\begin{equation}
\label{cp}
C_p=\int_0^1\mbox{d}s\cdot s\cdot\mbox{mmse}_s[\ln p(Y|X)|Y],
\end{equation}
where $\mbox{mmse}_s[\ln p(Y|X)|Y]$ is defined w.r.t.\ the joint pmf
\begin{equation}
q_s(x,y)=q(y)v_s(x|y)=q(y)\cdot\frac{p(x)w^s(y|x)}{Z_y(s)}.
\end{equation}
Eq.\ (\ref{cp}) seems to be less useful than the analogous rate--distortion formulas, 
for a very simple reason:
Since the channel is given, then once the input pmf $p$ is given 
too (which is required for the use of (\ref{cp})), one can simply compute
the mutual information, which is easier than applying (\ref{cp}). This is
different from the situation in the rate--distortion problem, where even if
both $p$ and $q$ are given, in order to compute $R_q(D)$ in the direct way, 
one still needs to minimize the mutual information
w.r.t.\ the channel between $X$ and $Y$. Eq.\ (\ref{cp}) is therefore
presented here merely for the purpose of drawing the duality.\\

\noindent
{\it Analogies With Statistical Mechanics}\\

As was shown in \cite{Rose94} and further advocated in \cite{Merhav08}, 
the Legendre relation (\ref{lgd1}) has a natural statistical--mechanical
interpretation, where $Z_x(s)$ plays the role of a partition function
of a system (indexed by $x$), $d(x,y)$ is an energy function (Hamiltonian)
and $s$ plays the role of inverse temperature (normally denoted by $\beta$
in the Physics literature). The minimizing $s$ is then the equilibrium inverse
temperature when $|\calX|$ systems (each indexed by $x$, with $n(x)=np(x)$
particles and Hamiltonian $\calE_x(y)=d(x,y)$) are brought into thermal
contact and a total energy of $nD$ is split among them. In this case,
$-R_q(D)$ is the thermodynamical entropy of the combined system and the
MMSE, which is $\mbox{d}D_s/\mbox{d}s$, is intimately related to the
heat capacity of the system.

An alternative, though similar, interpretation was given in
\cite{Merhav09a},\cite{Merhav09b},
where the parameter $s$ was interpreted 
as being proportional to a generalized force acting on the
system (e.g., pressure or magnetic field), and the distortion variable is
the conjugate physical quantity influenced by this force (e.g., volume in the case of
pressure, or magnetization in the case of a magnetic field). In this case, the
minimizing $s$ means the equal force that each one of the various subsystems is applying
on the others when they are brought into contact and they equilibrate (e.g.,
equal pressures between two volumes of a gas separated by piston which is free
to move). In this
case, $-R_q(D)$ is interpreted as the free energy of the system, and the
MMSE formulas are intimately related to the fluctuation--dissipation theorem
in statistical mechanics. 

More concretely, it was shown in \cite{Merhav09a} that given a source
distribution and a distortion measure, we can describe (at least
conceptually) a concrete physical system 
that emulates the rate--distortion
problem in the following manner:
When no force is applied to
the system, its total length is $nD_0$,
where $n$ is the number of particles in the system
(and also the block length in the rate--distortion problem),
and $D_0$ is as defined above.
If one applies to the system a contracting force, that increases
from zero to some final value
$\lambda$, such that the length of the system shrinks to
$nD$, where $D < D_0$ is
analogous to a prescribed distortion level,
then the following two facts hold true:
(i) An {\it achievable lower bound}
on the total amount of
mechanical work that must be carried out
by the contracting force in order to shrink the system to length $nD$, is
given by
\begin{equation}
W\ge nkTR_q(D),
\end{equation}
where $k$ is Boltzmann's constant and
$T$ is the temperature.
(ii) The final force
$\lambda$ is related to $D$ according to
$\lambda=kTR_q'(D)$, where $R_q'(\cdot)$ is the derivative of $R_q(\cdot)$.
Thus, the rate--distortion function plays the role of a fundamental limit,
not only in Information Theory, but in Physics as well.
 
\section{Examples}

In this section, we provide a few examples for the use of Theorem 1.
The first two examples are simple and well known, and their purpose
is just to demonstrate how to use 
this theorem in order to calculate rate--distortion
functions. The third example is aimed to demonstrate how Theorem 1
can be useful as a new method to 
evaluate the behavior of a certain rate--distortion
function (which is apparently not straightforward
to derive otherwise) at both the low distortion (a.k.a.\ high 
resolution) regime and the high distortion regime. Specifically,
we first derive, for this example, upper and lower bounds on $R(D)$, which
are applicable in certain ranges of high--distortion. 
These bounds have the same asymptotic behavior as $D$
tends to its maximum possible value, and so, they sandwich the 
exact high--distortion asymptotic behavior
of the true rate--distortion function. A similar analysis in then carried out in
the low distortion range, and again, the two bounds have the same limiting
behavior in the very low distortion limit.
In the fourth and last example, we show how Theorem 1
can easily be used to 
evaluate the high--resolution behavior of the rate distortion
function for a general power--law distortion measure of the form
$d(x,y)=|x-y|^r$.

\subsection{Binary Symmetric Source and Hamming Distortion}

Perhaps the simplest example is that of
the binary symmetric source (BSS) and the Hamming distortion
measure. In this case, the optimum $q$ is also symmetric. Here $\Delta=d(X,Y)$
is a binary RV with 
\begin{equation}
\mbox{Pr}\{\Delta=1|X-x\}=\frac{e^{-s}}{1+e^{-s}} 
\end{equation}
independently of
$x$. Thus, the MMSE estimator of $d(X,Y)$ based on $X$ is
\begin{equation}
\hat{\Delta}=\frac{e^{-s}}{1+e^{-s}},
\end{equation}
regardless of $X$,
and so the resulting MMSE (which 
is simply the variance in this case) is easily found to be
\begin{equation}
\mbox{mmse}_s(\Delta|X)=\frac{e^{-s}}{(1+e^{-s})^2}.
\end{equation}
Accordingly,
\begin{equation}
D=\frac{1}{2}-\int_0^s\frac{e^{-\hs}\mbox{d}
\hs}{(1+e^{-\hs})^2}=\frac{e^{-s}}{1+e^{-s}}
\end{equation}
and
\begin{eqnarray}
R(D)&=&\int_0^s\frac{\hs e^{-\hs}
\mbox{d}\hs}{(1+e^{-\hs})^2}\nonumber\\
&=&\ln 2 + \frac{se^{s}}{1+e^{s}}-\ln(1+e^{s})\nonumber\\
&=&\ln 2-h_2\left(\frac{e^{s}}{1+e^{s}}\right)\nonumber\\
&=&\ln 2-h_2(D),
\end{eqnarray}
where $h_2(u)=-u\ln u-(1-u)\ln(1-u)$ is the binary entropy function.

\subsection{Quadratic distortion and Gaussian Reproduction}

Another classic example concerns a general source with $\sigma_x^2= E\{X^2\} <
\infty$, the quadratic distortion $d(x,y)=(x-y)^2$, and a Gaussian reproduction
distribution, namely, $q(y)$ is the pdf of a zero--mean Gaussian RV with
variance $\sigma_y^2=\sigma_x^2-D$, for 
a given $D < \sigma_x^2$. In this case, it well known that
$R_q(D)=\frac{1}{2}\ln\frac{\sigma_x^2}{D}$ (even without assuming that
the source $X$ is Gaussian). 
We now demonstrate how this
result is obtained from the MMSE formula of Theorem 1.\footnote{We are not
arguing here that this is the simplest way to calculate $R_q(D)$ in this
example, the purpose is merely to demonstrate how Theorem 1 can be used.}

First, observe that since $q(y)$ is the pdf pertaining to 
$\calN(0,\sigma_x^2-D)$,
then
\begin{equation}
w_s(y|x)=\frac{q(y)e^{-s(y-x)^2}}{\int_{-\infty}^{+\infty}\mbox{d}y'q(y')e^{-s(y'-x)^2}}
\end{equation}
is easily found to correspond to the Gaussian additive channel
\begin{equation}
Y= \frac{2s(\sigma_x^2-D)}{1+2s(\sigma_x^2-D)}\cdot X+Z
\end{equation}
where $Z$ is a zero--mean Gaussian RV with variance
$\sigma_z^2=(\sigma_x^2-D)/[1+2s(\sigma_x^2-D)]$, and $Z$ is uncorrelated with $X$.
Now,
\begin{eqnarray}
\Delta&=&(Y-X)^2\nonumber\\
&=&\left[Y-\frac{2s(\sigma_x^2-D)}{1+2s(\sigma_x^2-D)}\cdot X
-\frac{X}{1+2s(\sigma_x^2-D)}\right]^2\nonumber\\
&=&(Z-\alpha X)^2\nonumber\\
&=&Z^2-2\alpha XZ+\alpha^2X^2
\end{eqnarray}
where $\alpha\dfn 1/[1+2s(\sigma_x^2-D)]$.
Thus, the MMSE estimator of $\Delta$ given $X$ is obtained by
\begin{eqnarray}
\hat{\Delta}&=&\bE\{\Delta|X\}\nonumber\\
&=&\bE\{Z^2|X\}-2\alpha X\bE\{Z|X\}+\alpha^2X^2\nonumber\\
&=&\bE\{Z^2\}-2\alpha X\bE\{Z\}+\alpha^2X^2\nonumber\\
&=&\bE\{Z^2\}+\alpha^2X^2\nonumber\\
&=&\sigma_z^2+\alpha^2X^2,
\end{eqnarray}
which yields
\begin{eqnarray}
& &\mbox{mmse}_s\{\Delta|X\}\nonumber\\
&=&
\bE\{(\hat{\Delta}-\Delta)^2\}\nonumber\\
&=&\bE\{(\sigma_z^2+\alpha^2X^2-Z^2+
2\alpha XZ-\alpha^2X^2)^2\}\nonumber\\
&=&2\sigma_z^4+4\alpha^2\sigma_x^2\sigma_z^2\nonumber\\
&=&\frac{2(\sigma_x^2-D)^2}{[1+2s(\sigma_x^2-D)]^2}+
\frac{4\sigma_x^2(\sigma_x^2-D)}{[1+2s(\sigma_x^2-D)]^3}.
\end{eqnarray}
Now, in our case, $D_0=\sigma_x^2+\sigma_y^2=2\sigma_x^2-D$, and so,
for $s=1/(2D)$, we get
\begin{eqnarray}
D_s&=&D_0-\int_0^s\mbox{d}\hs\cdot\mbox{mmse}_{\hs}(\Delta|X)\nonumber\\
&=&2\sigma_x^2-D-\nonumber\\
& &2(\sigma_x^2-D)^2\int_0^{1/2D}\frac{\mbox{d}\hs}{[1+2\hs(\sigma_x^2-D)]^2}-\nonumber\\
& &4\sigma_x^2(\sigma_x^2-D)\int_0^{1/2D}\frac{\mbox{d}\hs}{[1+2\hs(\sigma_x^2-D)]^3}\nonumber\\
&=&2\sigma_x^2-D+\nonumber\\
& &(\sigma_x^2-D)\left[\frac{1}{1+2s(\sigma_x^2-D)}\right]_0^{1/2D}+\nonumber\\
& &\sigma_x^2\left\{\frac{1}{[1+2s(\sigma_x^2-D)]^2}\right\}_0^{1/2D}
\end{eqnarray}
which, after some straightforward algebra, gives $D_s=D$. I.e.,
$s$ and $D$ are indeed related by $s=1/(2D)$, or $D=1/(2s)$.
Finally, 
\begin{eqnarray}
R_q(D)&=&\int_0^s\mbox{d}\hs\cdot\hs\cdot\mbox{mmse}_{\hs}(\Delta|X)\nonumber\\
&=&2(\sigma_x^2-D)^2\int_0^{1/2D}\frac{\hs\mbox{d}\hs}{[1+2\hs(\sigma_x^2-D)]^2}+\nonumber\\
& &4\sigma_x^2(\sigma_x^2-D)\int_0^{1/2D}
\frac{\hs\mbox{d}\hs}{[1+2\hs(\sigma_x^2-D)]^3}\nonumber\\
&=&\frac{1}{2}\left\{\ln[1+2s(\sigma_x^2-D)]+\right.\nonumber\\
& &\left.\frac{1}{1+2s(\sigma_x^2-D)}\right\}_0^{1/2D}+\nonumber\\
& &\frac{\sigma_x^2}{\sigma_x^2-D}
\left[\frac{1}{2[1+2s(\sigma_x^2-D)]^2}-\right.\nonumber\\
& &\left.\frac{1}{1+2s(\sigma_x^2-D)}\right]_0^{1/2D}
\end{eqnarray}
which yields, after a simple algebraic manipulation, 
$R_q(D)=\frac{1}{2}\ln\frac{\sigma_x^2}{D}$.

\subsection{Quadratic Distortion and Binary Reproduction}

In this example, we again assume the quadratic distortion measure, but now, instead
of Gaussian reproduction codewords, we impose binary reproduction, $y\in\{-a,+a\}$,
where $a$ is a given constant.\footnote{The derivation, in this example, can be
extended to apply also to larger finite reproduction alphabets.}
Clearly, if the pdf of the source $X$ is
symmetric about the origin, then the best output distribution is also
symmetric, i.e.,
$q(+a)=q(-a)=1/2$. Thus, $R_q(D)=R(D)$ for every $D$, given this choice of $q$.
The channel $w_s(y|x)$ is now given by
\begin{equation}
w_s(y|x)=\frac{e^{-s(y-x)^2}}{e^{-s(x-a)^2}+e^{-s(x+a)^2}}=\frac{e^{2sxy}}{2\cosh(2asx)}.
\end{equation}
Note that in this case, the minimum possible distortion 
(obtained for $s\to\infty$) is given by
$D_{\infty}=\bE\{[X-a\mbox{sgn}(X)]^2\}$. Thus, the rate--distortion function
is actually defined only for $D\ge D_{\infty}$. The maximum distortion of
interest is $D_0=\sigma_x^2+a^2$, pertaining to the choice $s=0$, where $X$ and $Y$ are
independent. 
To the best of our knowledge, there is no closed form expression for $R(D)$ in
this example. The parametric representation of $D_s$ and $R(D_s)$, both as
functions of $s$, does not seem to lend itself to an explicit formula of
$R(D)$. The reason is that
\begin{eqnarray}
D_s&=&\bE\{(Y-X)^2\}\nonumber\\
&=&\sigma_x^2+a^2-2\bE\{XY\}\nonumber\\
&=&\sigma_x^2+a^2-2\bE\{X\cdot\bE\{Y|X\}\}\nonumber\\
&=&\sigma_x^2+a^2-2a\bE\{X\tanh(2asX)\}
\end{eqnarray}
and there is no apparent closed--form expression of $s$ a function of $D$,
which can be substituted into the expression of $R(D_s)$.

Consider the MMSE estimator of $\Delta=(Y-X)^2=X^2+a^2-2XY$:
\begin{eqnarray}
\hat{\Delta}&=&\bE\{(Y-X)^2|X\}\nonumber\\
&=&X^2+a^2-2X\bE\{Y|X\}\nonumber\\
&=&X^2+a^2-2aX\tanh(2asX).
\end{eqnarray}
The MMSE is then
\begin{eqnarray}
\mbox{mmse}_s(\Delta|X)&=&\bE\{[2X(Y-a\tanh(2asX))]^2\}\nonumber\\
&=&4a^2[\sigma_x^2-\bE\{X^2\tanh^2(2asX)\}].
\end{eqnarray}
We first use this expression to obtain upper and lower bounds on $R(D)$
which are asymptotically exact in the 
range of high distortion levels (small $s$).
Subsequently, we do the same for the range of low distortion (large $s$).\\

\noindent
{\it High Distortion.}
Consider first the high distortion regime.
For small $s$, we can safely upper bound $\tanh^2(2asX)$ by $(2asX)^2$ and get
\begin{eqnarray}
\mbox{mmse}_s(\Delta|X)&\ge&
4a^2(\sigma_x^2-4a^2s^2\bE\{X^4\})\nonumber\\
&=&4a^2\sigma_x^2-16a^4\rho_x^4s^2
\end{eqnarray}
where $\rho_x^4\dfn E\{X^4\}$.
This results in the following lower bound to $R(D_s)$:
\begin{eqnarray}
R(D_s)&=&\int_0^s\mbox{d}\hs\cdot\hs\cdot\mbox{mmse}_{\hs}(\Delta|X)\nonumber\\
&\ge&\int_0^s\mbox{d}\hs\cdot\hs[4a^2\sigma_x^2-16a^4\rho_x^4\hs^2]\nonumber\\
&=&2a^2\sigma_x^2s^2-4a^4\rho_x^4s^4\dfn r(s).
\end{eqnarray}
To get a lower bound to $D_s$, we need an upper bound to the MMSE.
An obvious upper bound (which is tight for small $s$) is given by
$4a^2\sigma_x^2$, which yields:
\begin{eqnarray}
D_s&=&D_0-\int_0^s\mbox{d}\hs\cdot\mbox{mmse}_{\hs}(\Delta|X)\nonumber\\
&\ge&D_0-\int_0^s\mbox{d}\hs\cdot(4a^2\sigma_x^2)\nonumber\\
&=&D_0-4a^2\sigma_x^2s
\end{eqnarray}
or 
\begin{equation}
s\ge \frac{D_0-D_s}{4a^2\sigma_x^2}.
\end{equation}
Consider now the range $s\in[0, \sigma_x/(2a\rho_x^2)]$, which is 
the range where $r(s)$ is monotonically increasing as a function of $s$.
In this range, a lower bound on $s$ would yield a lower bound on $r(s)$,
and hence a lower bound to $R(D_s)$. Specifically, for $s\in[0,
\sigma_x/(2a\rho_x^2)]$, we get
\begin{eqnarray}
R(D_s)&\ge&r(s)\nonumber\\
&\ge&r\left(\frac{D_0-D_s}{4a^2\sigma_x^2}\right)\nonumber\\
&=&\frac{(D_0-D_s)^2}{8a^2\sigma_x^2}-\frac{\rho_x^4(D_0-D_s)^4}{64a^4\sigma_x^8}.
\end{eqnarray}
In other words, we obtain the lower bound
\begin{equation}
\label{lowerbound}
R(D)\ge
\frac{(D_0-D)^2}{8a^2\sigma_x^2}-\frac{\rho_x^4(D_0-D)^4}{64a^4\sigma_x^8}\dfn
R_L(D).
\end{equation}
for the range of distortions $D\in[D_0-2a\sigma_x^3/\rho_x^2,D_0]$.
It is obvious that, at least in some range of high distortion levels,\
this bound is better than the Shannon lower bound,
\begin{equation}
R_S(D)=h(X)-\frac{1}{2}\ln(2\pi e D),
\end{equation}
where $h(X)$ is the differential entropy of $X$. This can be seen right away
from the fact that $R_S(D)$ vanishes at $D=(2\pi
e)^{-1}e^{2h(X)}\le\sigma_x^2$, whereas the bound $R_L(D)$ of
(\ref{lowerbound}) vanishes at $D_0=\sigma_x^2+a^2$, which is strictly
larger. 

By applying the above--mentioned upper bound to the MMSE in the rate equation,
and the lower bound to the MMSE -- in the distortion equation, we can also get an upper
bound to $R(D)$ in the high--distortion range, in a similar manner.
Specifically,
\begin{equation}
R(D_s)\le\int_0^s\mbox{d}\hs\cdot\hs(4a^2\sigma_x^2)=2a^2\sigma_x^2s^2,
\end{equation}
and 
\begin{eqnarray}
D_s&\le&D_0-\int_0^s\mbox{d}\hs(4a^2\sigma_x^2-16a^4\rho_x^4\hs^2)\nonumber\\
&=&D_0-4a^2\sigma_x^2s+\frac{16}{3}a^4\rho_x^4s^3\dfn \delta(s).
\end{eqnarray}
Considering again the range $s\in[0, \sigma_x/(2a\rho_x^2)]$, where
$\delta(s)$ is monotonically decreasing, the inverse function $\delta^{-1}(D)$
is monotonically decreasing as well, and so an upper bound on $R(D)$ will be
obtained by substituting $\delta^{-1}(D)$ instead of $s$ in the bound on the
rate, i.e., $R(D)\le 2a^2\sigma_x^2[\delta^{-1}(D)]^2$. To obtain an explicit
expression for $\delta^{-1}(D)$, we need to solve a cubic equation in $s$ and
select the relevant solution among the three. Fortunately, since this cubic equation has no
quadratic term, the expression of the solution can be found
trigonometrically and it is relatively simple (see, e.g., \cite[p.\
9]{mathhandbook}): Specifically, the cubic equation $s^3+As+B=0$ has solutions of the
form $s=m\cos\theta$, where $m=2\sqrt{-A/3}$ and $\theta$ is any solution to
the equation $\cos(3\theta)=\frac{3B}{Am}$. In other words,
the three solutions to the above cubic equation are
$s_i=m\cos\theta_i$, where
\begin{equation}
\theta_i=\frac{1}{3}\cos^{-1}\left(\frac{3B}{Am}\right)+\frac{2\pi
(i-1)}{3},~~~~~~i=1,2,3,
\end{equation}
with $\cos^{-1}(t)$ being 
defined as the unique solution to the equation $\cos\alpha=t$
in the range $\alpha\in[0,\pi]$.
In our case,
\begin{equation}
A=-\frac{3\sigma_x^2}{4a^2\rho_x^4},~~~B=\frac{3(D_0-D)}{16a^4\rho_x^4},
\end{equation}
and so, the relevant solution for $s$ (i.e., the one that tends to
zero as $D\to D_0$),
which is $\delta^{-1}(D)$, is given by
\begin{eqnarray}
& &\delta^{-1}(D)\nonumber\\
&=&\frac{\sigma_x}{a\rho_x^2}
\cos\left[\frac{1}{3}\cos^{-1}\left(
\frac{3\rho_x^2(D-D_0)}{4a\sigma_x^3}\right)+\frac{4\pi}{3}\right]\nonumber\\
&=&\frac{\sigma_x}{a\rho_x^2}\cos\left[
\frac{1}{3}\left(\frac{\pi}{2}+\sin^{-1}\left(
\frac{3\rho_x^2(D_0-D)}{4a\sigma_x^3}\right)\right)+\frac{4\pi}{3}\right]
\nonumber\\
&=&\frac{\sigma_x}{a\rho_x^2}\sin\left[
\frac{1}{3}\sin^{-1}\left(
\frac{3\rho_x^2(D_0-D)}{4a\sigma_x^3}\right)\right],
\end{eqnarray}
where $\sin^{-1}(t)$ is defined as the unique solution to the equation
$\sin\alpha =t$ in the range $\alpha\in[-\pi/2,\pi/2]$.
This yields the upper bound
\begin{eqnarray}
R(D)&\le&\frac{2\sigma_x^4}{\rho_x^4}\sin^2\left[
\frac{1}{3}\sin^{-1}\left(
\frac{3\rho_x^2(D_0-D)}{4a\sigma_x^3}\right)\right]\nonumber\\
&\dfn&R_U(D).
\end{eqnarray}
for the range of distortions $D\in[D_0-4a\sigma_x^3/(3\rho_x^2),D_0]$.

For very small $s$, since the upper and the lower bound
to the MMSE asymptotically coincide (namely,
$\mbox{mmse}_s(\Delta|X)\approx 4a^2\sigma_x^2$), then both $R_U(D)$ and
$R_L(D)$ exhibit the same behavior near $D=D_0$, and hence so does the
true rate--distortion function, $R(D)$, which is
\begin{equation}
R(D)\approx \frac{(D_0-D)^2}{8a^2\sigma_x^2}
\end{equation}
or, stated more rigorously,
\begin{equation}
\lim_{D\uparrow D_0}\frac{R(D)}{(D_0-D)^2}=\frac{1}{8a^2\sigma_x^2}.
\end{equation}
Note that the high--distortion behavior of $R(D)$ depends on the
pdf of $X$ only via its second 
order moment $\sigma_x^2$. On the other hand, the upper and lower
bounds, $R_U(D)$ and
$R_L(D)$,
depend only on $\sigma_x^2$ and the fourth order moment, $\rho_x^4$.

\begin{figure}[h!t!b!]
\centering
\includegraphics[width=8.5cm, height=8.5cm]{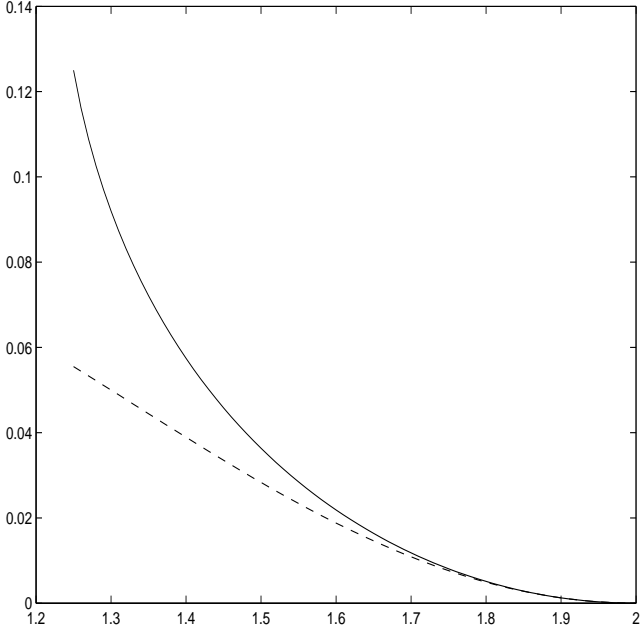}
\caption{The upper bound $R_U(D)$ (solid curve)
and the lower bound $R_L(D)$ (dashed curve) in the high--distortion regime
for $\sigma_x^2=a^2=1$ and $\rho_x^4=3$. The Shannon lower bound
vanishes in this distortion range.}
\label{bounds}
\end{figure}

In Fig.\ \ref{bounds}, we display the upper bound $R_U(D)$
(solid curve) and the lower bound $R_L(D)$ (dashed curve) for
the choice $\sigma_x^2=a^2=1$ 
(hence $D_0=\sigma_x^2+a^2=2$)
and $\rho_x^4=3$, which is suitable for
the Gaussian source. The range of displayed distortions, $[1.25,2]$,
is part of the range where both bounds are valid in this numerical example.
As can be seen, the 
functions $R_L(D)$ and $R_U(D)$ are very close 
throughout the interval $[1.7,2]$, which is a fairly
wide range of distortion levels. 
The corresponding Shannon lower bound,
in this case, which is $R_S(D)=\max\{0,\frac{1}{2}\ln\frac{1}{D}\}$,
vanishes for all $D\ge 1$ and hence also in the range displayed in the
graph.\\

\noindent
{\it Low Distortion.}
We now consider the small distortion regime, where $s$ is very large.
Define the function 
\begin{equation}
f(u)=\left(\frac{1-u}{1+u}\right)^2 ~~~u\in[0,1)
\end{equation}
and consider the Taylor series expansion of $f(u)$ around $u=0$,
which, for the sake of convenience, will be represented as
\begin{equation}
f(u)=1-\sum_{n=1}^\infty \phi_nu^n
\end{equation}
The coefficients $\{\phi_n\}$ will be determined explicitly in the sequel.
Now, clearly, $\tanh^2(2asx)\equiv f(e^{-4as|x|})$, and so we have
\begin{eqnarray}
& &\mbox{mmse}_s(\Delta|X)\nonumber\\
&=&4a^2\left[\sigma_x^2-\bE\{X^2f(\exp\{-4as|X|\})\}\right]\nonumber\\
&=&4a^2\left[\sigma_x^2-\bE\left\{X^2\left(1-\sum_{n=1}^\infty
\phi_ne^{-4ans|X|}\right)\right\}\right]\nonumber\\
&=&4a^2\sum_{n=1}^\infty\phi_n\bE\left\{X^2
e^{-4ans|X|}\right\}.
\end{eqnarray}
To continue from this point, 
we will have to let $X$ assume a certain pdf.
For convenience, let us select $X$ to have the Laplacian
pdf with parameter $\theta$, i.e.,
\begin{equation}
p(x) = \frac{\theta}{2}e^{-\theta|x|}.
\end{equation}
We then obtain
\begin{eqnarray}
\mbox{mmse}_s(\Delta|X)&=&2a^2\theta\sum_{n=1}^\infty\phi_n\int_{-\infty}^{+\infty}
x^2e^{-(\theta+4ans)|x|}\mbox{d}x\nonumber\\
&=&8a^2\theta\sum_{n=1}^\infty\frac{\phi_n}{(\theta+4ans)^3}.
\end{eqnarray}
Thus,
\begin{eqnarray}
& &R(D_s)\nonumber\\
&=&R(D_\infty)-\int_s^\infty
\mbox{d}\hs\cdot\hs\cdot\mbox{mmse}_s(\Delta|X)\nonumber\\
&=&1-8a^2\theta\sum_{n=1}^\infty\phi_n\cdot\int_s^\infty
\frac{\mbox{d}\hs\cdot\hs}{(\theta+4an\hs)^3}\nonumber\\
&=&1-\frac{\theta}{2}\sum_{n=1}^\infty\frac{\phi_n}{n^2}\left[
\frac{1}{\theta+4ans}-\frac{\theta}{2(\theta+4ans)^2}\right].
\end{eqnarray}
Thus far, our derivation has been exact. We now make an approximation that
applies for large $s$ by neglecting the terms proportional to
$(\theta+4ans)^{-2}$ and by neglecting $\theta$ compared to $4ans$ in the
denominators of $1/(\theta+4ans)$. This results in the approximation
\begin{equation}
R(D_s)\approx
\tilde{R}(D_s)\dfn 1-\frac{\theta}{8as}\sum_{n=1}^\infty\frac{\phi_n}{n^3}.
\end{equation}
Let us denote $C\dfn\frac{\theta}{8a}\sum_{n=1}^\infty\frac{\phi_n}{n^3}$.
Then, $\tilde{R}(D_s)=1-C/s$.
Applying a similar calculation to
$D_s=D_{\infty}+\int_s^\infty\mbox{d}\hs\cdot\mbox{mmse}_{hs}(\Delta|X)$,
yields, in a similar manner, the approximation
\begin{equation}
D_s\approx
\tilde{D}_s\dfn D_\infty+\frac{C}{2s^2}.
\end{equation}
It is easy now to express $s$ as a function of $D$ and substitute into the
rate equation to obtain
\begin{equation}
\label{highres}
R(D)\approx 1-\sqrt{2C(D-D_\infty)}.
\end{equation}
Finally, it remains to determine the coefficients $\{\phi_n\}$ and then the
constant $C$. The coefficients can easily be obtained by using the identity
$(1+u)^{-1}=\sum_{n=0}^\infty(-1)^nu^n$ ($u\in[0,1)$), 
which yields, after simple algebra, $\phi_n=4n(-1)^{n+1}$.
Thus,
\begin{equation}
C=\frac{\theta}{2a}\sum_{n=1}^\infty\frac{(-1)^{n+1}}{n^2}=\frac{\pi^2\theta}{24a}.
\end{equation}
and we have obtained a precise characterization of $R(D)$ in the
high--resolution regime:
\begin{equation}
\label{highreslim}
\lim_{D\downarrow
D_\infty}\frac{1-R(D)}{\sqrt{D-D_\infty}}=\sqrt{2C}=
\frac{\pi}{2}\cdot\sqrt{\frac{\theta}{3a}}.
\end{equation}
By applying a somewhat more refined analysis, one obtains 
(similarly as in the above derivation in the high distortion regime) upper and lower
bounds to $R(D_s)$ and $D_s$, this time, as polynomials in $1/s$.
These again lend themselves to the derivation of upper and lower bounds
on $R(D)$, which are applicable in certain intervals of low distortion.
Specifically, the resulting upper bound is
\begin{equation}
R(D)\le 1-\sqrt{2C(D-D_{\infty})}+C_1(D-D_{\infty}),
\end{equation}
where $C_1=\frac{9\theta}{\pi^2a}\sum_{n=1}^\infty\frac{(-1)^{n+1}}{n^3}$,
and it is valid in the range $D\in[D_\infty, D_{\infty}+C/(2C_1^2)]$.
The obtained lower bound is
\begin{equation}
R(D)\ge 1-\frac{\sqrt{6C(D-D_{\infty})}}{2\cos\left[\frac{1}{3}\sin^{-1}\left(
2C_1\sqrt{\frac{6(D-D_{\infty})}{C}}\right)+\frac{\pi}{6}\right]},
\end{equation}
and it applies to the range $D\in[D_\infty,D_\infty+C/(12C_1^2)]$.
Both bounds have the same leading term in asymptotic behavior,
which supports eq.\ (\ref{highreslim}).
The details of this derivation are omitted since they are very similar to
those of the high--distortion analysis.

\subsection{High Resolution for a General $L^r$ Distortion Measure}

Consider the case where the distortion measure is given by the $L^r$ metric,
$d(x,y)=|x-y|^r$ for some fixed $r > 0$. Let the reproduction symbols
be selected independently at random according to the uniform pdf
\begin{equation}
q(y)=\left\{\begin{array}{ll}
\frac{1}{2A} & |y|\le A\\
0 & \mbox{elsewhere}\end{array}\right.
\end{equation}
Then
\begin{equation}
w_s(y|x)=\frac{e^{-s|y-x|^r}}{\int_{-A}^{+A}\mbox{d}y'\cdot e^{-s|y'-x|^r}}
\end{equation}
and so
\begin{eqnarray}
D_s&=&\int_{-\infty}^{+\infty}\mbox{d}xp(x)\cdot
\frac{\int_{-A}^{+A}\mbox{d}y\cdot |x-y|^re^{-s|y-x|^r}}
{\int_{-A}^{+A}\mbox{d}y\cdot e^{-s|y-x|^r}}\nonumber\\
&=&-\int_{-\infty}^{+\infty}\mbox{d}xp(x)\cdot\frac{\partial}{\partial
s}\ln\left[\int_{-A}^{+A}\mbox{d}y\cdot e^{-s|y-x|^r}\right].
\end{eqnarray}
Now, in the high--resolution limit, where $s$ is very large, the
integrand $e^{-s|y-x|^r}$ decays very rapidly as $y$ takes values
away from $x$, and so, for every $x\in(-A,+A)$ (which for large
enough $A$, is the dominant interval for  
the outer integral over $p(x)\mbox{d}x$), the boundaries,
$-A$ and $+A$, of the inner integral can be extended to $-\infty$ and $+\infty$
within a negligible error term
(whose derivative w.r.t. $s$ 
is negligible too). Having done this, the inner integral no longer
depends on $\bx$, which also means that the outer integration over $x$ becomes
superfluous.
This results in
\begin{eqnarray}
D_s&=&-\frac{\partial}{\partial s}\ln\left[\int_{-\infty}^{+\infty}
\mbox{d}y\cdot e^{-s|y|^r}\right]\nonumber\\
&=&-\frac{\partial}{\partial s}\ln\left[s^{-1/r}\int_{-\infty}^{+\infty}
\mbox{d}(s^{1/r}y)e^{-|s^{1/r}y|^r}\right]\nonumber\\
&=&-\frac{\partial}{\partial s}\ln\left[s^{-1/r}\int_{-\infty}^{+\infty}
\mbox{d}t\cdot e^{-|t|^r}\right]\nonumber\\
&=&-\frac{\partial}{\partial s}\ln(s^{-1/r})\nonumber\\
&=&\frac{1}{rs}.
\end{eqnarray}
Thus,
\begin{equation}
\mbox{mmse}_s(\Delta|X)=-\frac{\mbox{d}D_s}{\mbox{d}s}=\frac{1}{rs^2},
\end{equation}
which yields
\begin{equation}
\frac{\mbox{d}R_q(D_s)}{\mbox{d}s}=s\cdot\mbox{mmse}_s(\Delta|X)=\frac{1}{rs}
\end{equation}
and so
\begin{eqnarray}
R_q(D_s)&=&K+\frac{1}{r}\ln s\nonumber\\
&=&K+\frac{1}{r}\ln\left(\frac{1}{rD_s}\right)
\end{eqnarray}
where $K$ is an integration constant. We have therefore
obtained that in the high--resolution limit, the
rate--distortion function w.r.t.\ $q$ behaves according to
\begin{equation}
R_q(D)=K'-\frac{1}{r}\ln D.
\end{equation}
with $K'=K-(\ln r)/r$. While this simple derivation does not determine yet the
constant $K'$, it does provide the correct characteristics 
of the dependence of $R_q(D)$ upon $D$
for small $D$. For the case of quadratic distortion, where $r=2$, one easily
identifies the familiar factor of $1/2$ in front of the log--distortion term.

The exact constant $K$ (or $K'$) can be determined by returning to the original expression of
$R_q(D)$ as the Legendre transform of the log--moment generating function of
the distortion (eq.\ (\ref{lgd1}), 
and setting there $s=1/(rD)$ as the minimizing $s$ for the given $D$.
The resulting expression turns out to be
\begin{equation}
K'=\ln\left[\frac{rA}{\Gamma(1/r)}\right]-\frac{1}{r}\ln(er).
\end{equation}

\section{Conclusion}

In this paper, we derived relations between the rate--distortion
function $R_q(D)$ and the MMSE in estimating the distortion given the source
symbol. These relations have been discussed from several aspects,
and it was demonstrated how they can be used to obtain upper and
lower bounds on $R_q(D)$, as well as the exact asymptotic behavior in very
high and very low distortion. 

The bounds derived in our examples were
induced from purely mathematical bounds on the expression of the MMSE
directly. We have not explored, however, examples of bounds on $R_q(D)$
that stem from estimation--theoretic bounds on the MMSE, as was described in
Section III. In future work, it would be interesting to explore the usefulness
of such bounds as well. Another interesting direction for further work would be
to make an attempt to extend our results to rate--distortion functions 
pertaining to more involved settings, such as successive refinement coding,
and situations that include side information.

\section*{Appendix}

\noindent
{\it Proof of Theorem 1.}\\
Consider a random selection of a codebook of $M=e^{nR}$ codewords, where the
various codewords are drawn independently, and each codeword,
$\bY=(Y_1,\ldots,Y_n)$, is drawn according to the product measure
$Q(\by)=\prod_{i=1}^nq(y_i)$. Let $\bx=(x_1,\ldots,x_n)$ be a typical source
vector, i.e., the number of times each symbol $x\in\calX$ appears in
$\bx$ is (very close to) $np(x)$. We now ask what
is the probability of the event
$\{\sum_{i=1}^nd(x_i,Y_i)\le nD\}$? As this is a large deviations event
whenever $D < \sum_{x,y}p(x)q(y)d(x,y)$, this probability must decay
exponentially
with some rate function $I_q(D) > 0$, i.e.,
\begin{equation}
I_q(D)=\lim_{n\to\infty}\left[-\frac{1}{n}\ln\mbox{Pr}\left\{\sum_{i=1}^nd(x_i,Y_i)\le
nD\right\}\right].
\end{equation}
The function $I_q(D)$ can be determined in two ways.
The first is by the method of types \cite{CK81}, which easily yields
\begin{equation}
\label{mot}
I_q(D)=\min[I(X;Y')+D(q'\|q)],
\end{equation}
where the $Y'$ is an auxiliary random variable governed by
$q'(y)=\sum_{x\in\calX}p(x)w(y|x)$ and the minimum is over all
conditional pmf's $\{w(y|x)\}$ that satisfy the inequality
$\sum_{x\in\calX}p(x)\sum_{y\in\calY}w(y|x)d(x,y)\le D$.
The second method is based on large deviations theory \cite{DZ93} (see also
\cite{Merhav08}), which yields
\begin{equation}
\label{ldt}
I_q(D)=-\min_{s\ge 0}\left[sD+\sum_{x\in\calX}p(x)\ln Z_x(s)\right].
\end{equation}
We first argue that $I_q(D)=R_q(D)$. 
The inequality $I_q(D)\le R_q(D)$ is
obvious, as $R_q(D)$ is obtained by confining the minimization over the
channels in (\ref{mot}) so as to comply with the additional constraint that
$\sum_{x\in\calX}p(x)w(y|x)=q(y)$ for all $y\in\calY$. 
The reversed
inequality,
$I_q(D)\ge R_q(D)$, is obtained by the following coding argument:
On the one hand, a trivial extension of the converse to the rate--distortion
coding theorem \cite[p.\ 317]{CT06}, shows that
$R_q(D)$ is a lower bound to the rate--distortion performance
of any code that satisfies $\frac{1}{n}\sum_{i=1}^n\mbox{Pr}\{Y_i=y\}=q(y)$ for all
$y\in\calY$.\footnote{To see why this is true, 
consider the functions $\delta_{k}(y)$, $y,k\in\calY$ (each
of which
is defined as equal one for
$y=k$ and zero otherwise) as $|\calY|$ distortion measures, indexed by
$k\in\calY$, 
and consider the rate--distortion function w.r.t.\ the usual distortion constraint
and the $|\calY|$ additional
``distortion constraints'' $\bE\{\delta_k(Y)\}\le q(k)$ for all $k\in\calY$,
which, when satisfied, they all must be achieved with equality (since they must sum
to unity). The
rate--distortion function w.r.t.\ these $|\calY|+1$ constraints, which is
exactly $R_q(D)$, is easily shown (using the standard method) to be jointly convex in $D$ and $q$.}
On the other hand, we next show that $I_q(D)$ is an achievable rate
for codes in this class. 

Consider the the random coding mechanism described in
the first paragraph of this proof, with $R=I_q(D)+\epsilon$, 
with $\epsilon >0$
being arbitrarily small. Since the
probability that for a single randomly drawn codeword,
$\mbox{Pr}\{\sum_{i=1}^nd(x_i,Y_i)\le nD\}$ is of the exponential order of
$e^{-nI_q(D)}$, then the random selection of a codebook
of size $e^{n[I_q(D)+\epsilon]}$ constitutes
$e^{n[I_q(D)+\epsilon]}$ independent trials of an experiment whose probability
of success is of the exponential order of $e^{-nI_q(D)}$. Using standard random coding
arguments, the probability that at least one codeword, in that codebook, would fall within
distance $nD$ from the given typical $\bx$ becomes overwhelmingly large as
$n\to\infty$. Since
this randomly selected codebook satisfies also $\frac{1}{n}\sum_{i=1}^n\mbox{Pr}\{Y_i=y\}\to q(y)$
in probability (as $n\to\infty$)
for all $y\in\calY$ (by the weak law of large numbers), then $I_q(D)$ is
an achievable rate within the class of codes that satisfy
$\frac{1}{n}\sum_{i=1}^n\mbox{Pr}\{Y_i=y\}\to q(y)$ for all $i$. 

Thus, $I_q(D)\ge R_q(D)$, which together with the reversed inequality proved
above, yields the equality $I_q(D)=R_q(D)$. Consequently, according to eq.\ (\ref{ldt}),
we have established the relation\footnote{
Eq.\ (\ref{lgdr}) appears also in \cite[p.\ 90, Corollary 4.2.3]{Gray90},
with a completely different proof, for the special case where $q$ minimizes
both sides of the equation (and hence it refers to $R(D)$). However, the
extension of that proof to a generic $q$ is not apparent to be straightforward because
here the minimization over the channels is limited by the reproduction
distribution constraint.}
\begin{equation}
\label{lgdr}
R_q(D)=-\min_{s\ge 0}\left[sD+\sum_{x\in\calX}p(x)\ln Z_x(s)\right].
\end{equation}
As this minimization problem is a convex problem ($\ln Z_x(s)$ is convex in
$s$), the minimizing $s$ for a given $D$ is obtained by taking the derivative
of the r.h.s., which leads to
\begin{eqnarray}
D&=&-\sum_{x\in\calX}p(x)\cdot\frac{\partial\ln Z_x(s)}{\partial s}\nonumber\\
&=&\sum_{x\in\calX}p(x)\cdot\frac{\sum_{y\in\calY}q(y)d(x,y)e^{-sd(x,y)}}
{\sum_{y\in\calY}q(y)e^{-sd(x,y)}}.
\end{eqnarray}
This equation yields the distortion level $D$ for a given value of the
minimizing $s$ in eq.\ (\ref{lgdr}). Let us then denote 
\begin{equation}
\label{ds}
D_s=\sum_{x\in\calX}p(x)\cdot\frac{\sum_{y\in\calY}q(y)d(x,y)e^{-sd(x,y)}}
{\sum_{y\in\calY}q(y)e^{-sd(x,y)}}.
\end{equation}
This notation obviously means that
\begin{equation}
\label{rds}
R_q(D_s)=-sD_s-\sum_{x\in\calX}p(x)\ln Z_x(s).
\end{equation}
Taking the derivative of (\ref{ds}), we readily obtain
\begin{eqnarray}
\frac{\mbox{d}D_s}{\mbox{d}s}&=&\sum_{x\in\calX}p(x)\frac{\partial}{\partial
s}\left[\frac{\sum_{y\in\calY}q(y)d(x,y)e^{-sd(x,y)}}
{\sum_{y\in\calY}q(y)e^{-sd(x,y)}}\right]\nonumber\\
&=&-\sum_{x\in\calX}p(x)
\left[\frac{\sum_{y\in\calY}q(y)d^2(x,y)e^{-sd(x,y)}}{\sum_{y\in\calY}q(y)e^{-sd(x,y)}}-\right.\nonumber\\
& &\left.\left(\frac{\sum_{y\in\calY}q(y)d(x,y)e^{-sd(x,y)}}
{\sum_{y\in\calY}q(y)e^{-sd(x,y)}}\right)^2\right]\nonumber\\
&=&-\sum_{x\in\calX}p(x)\cdot\mbox{Var}_s\{d(x,Y)|X=x\}\nonumber\\
&=&-\mbox{mmse}_s(\Delta|X),
\end{eqnarray}
where $\mbox{Var}_s\{d(x,Y)|X=x\}$ is the variance of $d(x,Y)$ w.r.t.\
the conditional pmf $\{w_s(y|x)\}$. The last line follows from the fact the
expectation of $\mbox{Var}_s\{d(X,Y)|X\}$ w.r.t.\ $\{p(x)\}$ is exactly the
MMSE of $d(X,Y)$ based on $X$.
The integral forms of this equation
are then precisely as in part (a) of the theorem with the corresponding integration
constants. Finally, differentiating both sides of eq.\ (\ref{rds}), we get
\begin{eqnarray}
\frac{\mbox{d}R(D_s)}{\mbox{d}s}&=&-s\cdot\frac{\mbox{d}D_s}{\mbox{d}s}-D_s-
\sum_{x\in\calX}p(x)\cdot\frac{\partial\ln Z_x(s)}{\partial s}\nonumber\\
&=&-s\cdot\frac{\mbox{d}D_s}{\mbox{d}s}-D_s+D_s\nonumber\\
&=&-s\cdot\frac{\mbox{d}D_s}{\mbox{d}s}\nonumber\\
&=&s\cdot\mbox{mmse}_s(\Delta|X),
\end{eqnarray}
which when integrated back, yields part (b) of the theorem.
This completes the proof of Theorem \ref{thm1}.


\begin{thebibliography}{1}

\bibitem{Berger71}
T.~Berger, {\it Rate Distortion Theory: A Mathematical Basis for Data
Compression}, Prentice--Hall, Englewood Cliffs, New Jersey, U.S.A., 1971.

\bibitem{CT06}
T.~M.~Cover and J.~A.~Thomas, {\it Elements of Information Theory}, (second
edition),
John Wiley \& Sons, Inc., New York, 2005.

\bibitem{CK81}
I. Csisz\'ar and J.
K\"orner, {\sl Information Theory: Coding Theorems for Discrete
Memoryless Systems.\/} New York: Academic, 1981.

\bibitem{DZ93}
A.~Dembo and O.~Zeitouni, {\it Large Deviations Techniques and Applications},
John and Bartlett Publishers, 1993.

\bibitem{Gray90}
R.~M.~Gray, {\it Source Coding Theory}, Kluwer Academic Publishers, 1990.

\bibitem{GSV05}
D.~Guo, S.~Shamai (Shitz), and S.~Verd\'u, ``Mutual information and minimum
mean--square error in Gaussian channels,'' {\it IEEE Trans.\ Inform.\ Theory}, vol.\ 51,
no.\ 4, pp.\ 1261--1282, April 2005.

\bibitem{mathhandbook}
M.~Fogiel, {\it Handbook of Mathematical, Scientific, and Engineering
Formulas, Tables, Functions, Graphs, Transforms}, Research and Education
Association, Piscataway, New Jersey, U.S.A., 1997.

\bibitem{Merhav08}
N.~Merhav,
``An identity of Chernoff bounds
with an interpretation in statistical physics and
applications in information theory,''
{\it IEEE Trans.\ Inform.\ Theory}, vol.\ 54, no.\ 8, pp.\
3710--3721, August 2008.

\bibitem{Merhav09a}
N.~Merhav, ``Another look at the physics of
large deviations with application to rate--distortion
theory,''\\
http://arxiv.org/PS\_cache/arxiv/pdf/0908/0908.3562v1.pdf.

\bibitem{Merhav09b}
N.~Merhav, ``On the physics of rate--distortion theory,'' to appear in 
{\it Proc.\ ISIT 2010}, Austin, Texas, U.S.A., June 2010.

\bibitem{Rose94}
K.~Rose, ``A mapping approach to rate-distortion computation and 
analysis,'' {\em IEEE Trans.~Inform.~Theory\/}, vol.\ 40, no.\ 6, pp.\
1939--1952, November 1994.

\bibitem{WW85}
E.~Weinstein and A.~J.~Weiss, ``Lower bounds on the mean square estimation
error,'' {\it Proc.\ of the IEEE}, vol.\ 73, no.\ 9, pp.\ 1433--1434,
September 1985.

\bibitem{Weiss85}
A.~J.~Weiss, {\it Fundamental Bounds in Parameter Estimation},
Ph.D.\ dissertation, Tel Aviv University, Tel Aviv, Israel, June 1985.

\bibitem{WZ71}
A.~D.~Wyner and J.~Ziv, ``Bounds on the rate--distortion function for
stationary sources with memory,'' 
{\em IEEE Trans.~Inform.~Theory\/}, vol.\ 17, no.\ 5, pp.\
508--513, September 1971.

\end{thebibliography}
\end{document}